\documentclass[pra, showpacs, twocolumn, floatfix]{revtex4}
\usepackage{graphicx}
\usepackage{epstopdf}
\usepackage{amsmath, amsfonts, amssymb, bm}
\begin{document}
\title{Long-time correlated quantum dynamics of phonon cooling}
\author{Sergiu \surname{Carlig}}

\author{Mihai A. \surname{Macovei}}
\email{macovei@phys.asm.md}
\affiliation{Institute of Applied Physics, Academy of Sciences of Moldova, 
Academiei str. 5, MD-2028 Chi\c{s}in\u{a}u, Moldova}
\date{\today}
\begin{abstract}
We investigate the steady-state cooling dynamics of vibrational degrees of freedom related to a nanomechanical oscillator coupled with a laser-pumped quantum 
dot in an optical resonator. Correlations between phonon-cooling and quantum-dot photon emission processes occur respectively when a photon laser 
absorption together with a vibrational phonon absorption is followed by photon emission in the optical resonator. Therefore, the detection of photons 
generated in the cavity mode concomitantly contribute to phonon cooling detection of the nanomechanical resonator.
\end{abstract}
\maketitle

\section{Introduction}
The nanomechanical resonator (NMR) is a relevant tool for building  ultra-sensitive measurement devises \cite{gr_cool, rew_nm,rew_m}. Therefore, its 
properties were and are continuously investigated. Outstanding works towards NMR cooling to quantum regimes were already reported 
\cite{cool,cool1,ccooll,ccool,cool2,cool3,cool4}. An important issue here is how to detect experimentally the mechanical vibrations of the NMR. One option 
is the superconducting quantum interferometer where the vibrations  of the NMR are detected via variation of the magnetic field \cite{interf}. 
The mechanical vibrations can be detected as well via a single-electron transistor which is extremely sensitive to electrical charges \cite{tranz}.  
Additionally one can use interference effects among the incident light on the NMR and the reflected one \cite{tabl_int}. Furthermore, 
high-sensitivity optical monitoring of a micro-mechanical resonator with a quantum-limited opto-mechanical sensor was reported in Ref.~\cite{opt_sens}, 
while fast sensitive displacement measurements in Ref.~\cite{fast}, respectively. Remarkably, the quantum motion of a nanomechanical resonator 
was experimentally observed in \cite{qmot}. The possibility of real time displacement detection by the luminescence signal and of displacement fluctuations 
by the degree of second-order coherence function was recently demonstrated in \cite{g2}.

\begin{figure}[b]
\centering
\includegraphics[height=3.5cm]{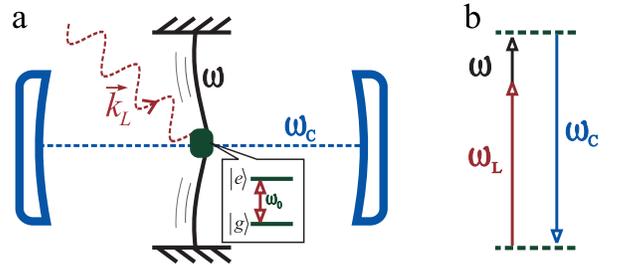}
\caption{\label{fig-1} 
(color online) Schematic model: (a) A laser-pumped two-level quantum dot with transition frequency $\omega_{0}$ is fixed on a nanomechanical resonator 
vibrating at frequency $\omega$. The quantum dot is interacting also with the quantized resonator optical mode of frequency $\omega_{C}$. 
(b) Correlated cooling dynamics occurs when a laser photon absorption together with a vibrational phonon absorption is accompanied by the emission of a cavity photon.}
\end{figure}

Here, we look for a regime where cooling of a nano-mecanical oscillator is correlated with emission processes such that the maximum photon detection 
corresponds to vibrational phonon minimum. For this, we investigate a laser-pumped two-level quantum dot which is fixed on a nanomechanical beam 
while suspended in an optical resonator (see Fig.~\ref{fig-1}a). If the quantum dot dynamics is faster than that of nano-beam and cavity dynamics, 
respectively, one arrives at a situation where laser photon and phonon absorption processes are accompanied with photon emission in the cavity mode 
(see Fig.~\ref{fig-1}b). Therefore, the cavity photon detection assures the cooling of the nanomechanical resonator. 

The article is organized as follows. In Sec. II we describe the theoretical framework used to obtain the master equation characterizing the correlated 
cooling dynamics of nanomechanical degrees of freedom. Section III deals with the corresponding equations of motion and discussion of the obtained 
results while the Summary is given in the last section, i.e., Sec. IV.

\section{Theoretical framework}
Let's consider a setup represented in Figure~({\ref{fig-1}}a): Inside an optical resonator is placed a NMR incorporating a laser pumped two-level quantum 
dot. The laser beam wave-vector is $\vec k_{L}$ while its frequency is $\omega_{L}$. The frequency of the optical resonator is $\omega_{C}$ and 
the nanomechanical vibrational frequency is $\omega$. The energy separation between the excited bare-state $|e\rangle$ and the ground one, 
$|g\rangle$, is denoted by $\hbar\omega_{0}$.  The master equation describing the whole system in the Born-Markov approximations and in a frame 
rotating at the laser frequency $\omega_{L}$ is: 
\begin{eqnarray}
&&\frac{d}{dt}\rho(t) + \frac{i}{\hbar}[H,\rho]=-\gamma [S^{+},S^{-}\rho] - \gamma_{c}[S_{z},S_{z}\rho] \nonumber \\
&&- \kappa_{a}[a^{\dagger},a\rho] - \kappa_{b}(1+\bar{n})[b^{\dagger},b\rho] - \kappa_{b}\bar{n}[b,b^{\dagger}\rho] + H.c., \nonumber \\
\label{Me}
\end{eqnarray}
where $S_{z}$ and $S^{\pm}$ are the qubit operators satisfying the standard commutation relations, while $\{a^{\dagger},a\}$ and 
$\{ b^{\dagger},b\}$ are the generation and annihilation operators for photon and phonon subsystems, respectively, and obey the boson 
commutation relations  \cite{kmek}. $\gamma$ and $\gamma_c$ are the single-qubit spontaneous decay and dephasing rates, respectively, 
whereas $\kappa_{a}(\kappa_{b})$ is the photon (phonon) resonator damping rate, and $\bar n$ is the mean-phonon number corresponding 
to temperature $T$ and vibrational frequency $\omega$. The Hamilton operator, i.e. $H$, is given by the following expression:
\begin{eqnarray}
H&=& \hbar \Delta S_{z} - \hbar \Delta_{1}a^{\dagger}a + \hbar \omega b^{\dagger}b + \hbar \Omega(S^{+}+ S^{-}) \nonumber \\
 &+&\hbar g(a^{\dagger}S^{-} +  aS^{+}) + \hbar\lambda S_{z} (b^{\dagger}+b). \label{Hmn} 
\end{eqnarray}
In Eq.~(\ref{Hmn}), the first three terms describe the free energies of the artificial two-level system as well as of the optical and 
mechanical modes. The fourth and the fifth terms characterize the interaction of the quantum dot with the laser field and optical resonator 
mode, respectively. The last term takes into account the interaction of the vibrational degrees of freedom with the radiator \cite{cool}. 
Correspondingly, $g$ and $\lambda$ denote the interaction strengths among the two-level emitter and the involved optical and mechanical 
modes, while $\Omega$ is the corresponding Rabi frequency due to external laser pumping. $\Delta=\omega_{0}-\omega_{L}$ describes
the detuning of the laser frequency from the two-level transition frequency, while $\Delta_{1}=\omega_{L} - \omega_C$  accordingly is the 
detuning of the cavity frequency from the laser one.

For our purpose, it is more appropriate to use the laser-qubit dressed-state representation given by \cite{SCMM}: $|g\rangle=\sin{\theta}| 
+ \rangle + \cos{\theta}|-\rangle$ and $|e\rangle=\cos{\theta}|+\rangle - \sin{\theta}|-\rangle$, with $|+\rangle$ and $|-\rangle$ being 
the corresponding states in the dressed-state picture. Here, $2\theta$ is the angle in the right triangle, drawn in imaginative space of 
frequencies, with adjoining cathetus $\Delta/2$ and opposite cathetus $\Omega$, and, therefore, $\cot{2\theta}=\Delta/2\Omega$. In the 
case when $\Omega \gg \{\gamma,\gamma_{c}\} \gg \kappa_{a,b}$ while $\Omega \gg \{g,\lambda\} >\{\gamma,\gamma_{c}\}$, meaning 
that the dynamics of the cavity photon and NMR phonon subsystems are slower than the quantum dot dynamics, one can eliminated the 
quantum dot variables (see also \cite{SCMM,zb1,kzek,gxl}). Thus, the master equation describing the cavity and NMR degrees of freedom can be 
represented as:
\begin{eqnarray}
\frac{d}{dt}\rho(t) &+&\frac{i}{2}(\Delta_{1}+\omega)[b^{\dagger}b-a^{\dagger}a,\rho] = \nonumber \\
&-& A^{\ast}_{1}[a,a^{\dagger}\rho] - B^{\ast}_{1}[a^{\dagger},a\rho] - A^{\ast}_{2}[b,b^{\dagger}\rho] \nonumber \\
&-& B^{\ast}_{2}[b^{\dagger},b\rho]+ C^{\ast}_{1}[b,a^{\dagger}\rho] + D^{\ast}_{1}[b^{\dagger},a\rho] 
\nonumber \\
&+& C^{\ast}_{2}[a^{\dagger},b\rho] + D^{\ast}_{2}[a,b^{\dagger}\rho] + H.c.. \label{MeCN}
\end{eqnarray}
Here $"\ast"$ means complex conjugation, whereas
\begin{eqnarray*}
A^{\ast}_{1}&=&\frac{1}{4}\frac{g^2 \sin ^2{2 \theta}}{\Gamma_{\shortparallel} - i\Delta_{1} }+\frac{g^2 P_{-} \sin^4{\theta}}{\Gamma_{\perp} 
+i (2\Omega_{R} -\Delta_{1} )}\nonumber\\
&+&\frac{g^2 P_{+}\cos ^4{\theta}}{\Gamma_{\perp}- i (2 \Omega_{R}+\Delta_{1}  )}, 
\end{eqnarray*}
\begin{eqnarray*}
A^{\ast}_{2}&=&\frac{1}{4}\bigg(\frac{\lambda^{2}\cos^{2}{2\theta}}{\Gamma_{\shortparallel}+i \omega }+\frac{\lambda^{2}P_{-}
\sin^{2}{2\theta}}{\Gamma_{\perp} + i(2 \Omega_{R} +\omega )}\nonumber\\
&+&\frac{\lambda^{2}P_{+}\sin^{2}{2 \theta}}{\Gamma_{\perp} - i(2 \Omega_{R}-\omega  )}\bigg)+\kappa_{b}\bar{n}, \nonumber \\
C^{\ast}_{1}&=&\frac{P_{+}}{2}\frac{g\lambda \sin{2\theta}\cos^{2}{\theta}}{\Gamma_{\perp} -i (2\Omega_{R}+\Delta_{1}) }
-\frac{P_{-}}{2}\frac{g\lambda \sin{2\theta}\sin^{2}{\theta}}{\Gamma_{\perp}+ i (2\Omega_{R}-\Delta_{1})} \nonumber\\
&-&\frac{1}{4}\frac{g\lambda \sin{2 \theta}\cos{2\theta}}{\Gamma_{\shortparallel} -i\Delta_{1}}, \nonumber\\
C^{\ast}_{2}&=&\frac{P_{-}}{2}\frac{g\lambda \sin{2\theta}\cos^{2}{\theta}}{\Gamma_{\perp} + i(2\Omega_{R}-\omega) }
-\frac{P_{+}}{2}\frac{g\lambda \sin{2\theta}\sin^{2}{\theta}}{\Gamma_{\perp}- i (2\Omega_{R}+\omega)}\nonumber\\
&-&\frac{1}{4}\frac{g\lambda \sin{2\theta}\cos{2\theta}}{\Gamma_{\shortparallel} - i\omega }.
\end{eqnarray*}
Other parameters are: $\Omega_{R}=\sqrt{\Omega^{2}+(\Delta/2)^{2}}$, $\Gamma_{\shortparallel}$=$\gamma(1-\cos^{2}{2\theta})+
\gamma_{c}\sin^{2}{2\theta}$, 
$\Gamma_{\perp}$=$4\gamma_0+\gamma_{+}+\gamma_{-}$, $\gamma_{+}$=$\gamma \cos^4{\theta} + \frac{\gamma_{c}}{4}\sin^{2}{2\theta}$, 
$\gamma_{-}$=$\gamma \sin^{4}{\theta} + \frac{\gamma_{c}}{4}\sin^{2}{2\theta}$ and $\gamma_{0}$=$\frac{1}{4}(\gamma \sin^2{2\theta}
+\gamma_{c} \cos^2{2\theta})$. The dressed-state populations are given by:
\begin{eqnarray*}
P_{+} = \frac{\gamma_-}{\gamma_++\gamma_-}, ~~{\rm and}~~ P_{-}=\frac{\gamma_+}{\gamma_++\gamma_-}.
\end{eqnarray*}
In Eq.~(\ref{MeCN})$, B^{\ast}_{i}$ can be obtained from $A^{\ast}_{i}$ via $P_{\mp} \leftrightarrow P_{\pm}$ as well as by adding $\kappa_{a}$ to 
$B^{\ast}_{1}$ and $\kappa_{b}$ to $B^{\ast}_{2}$, correspondingly. Respectively, $D^{\ast}_{i}$ can be obtained from $C^{\ast}_{i}$ through 
$P_{\mp} \leftrightarrow P_{\pm}$, and $\{i \in 1,2\}$. Notice that in obtaining Eq.~(\ref{MeCN}) we have ignored rapidly oscillating terms at frequencies: 
$\pm 2\Delta_{1}, \pm (\Delta_{1}- \omega)$ and $\pm 2\omega$, that is, we are interested in a situation where $\omega \approx -\Delta_{1}$. 

\section{Results and discussion}
In the following, we shall describe the steady-state correlated cooling dynamics of the vibrational degrees of freedom. Actually, adjusting the laser 
frequency to fulfill the condition $\omega + \Delta_{1} \approx 0$, one shall look at situations when simultaneously photon laser and vibrational phonon 
absorption processes are accompanied by a photon emission in the cavity mode, see Figure~(\ref{fig-1}b).  Using the master equation (\ref{MeCN}), the 
following equations of motion can be obtained for the mean photon and phonon numbers, respectively:
\begin{eqnarray}
\frac{d}{dt}\langle a^{\dagger}a\rangle& = &\langle a^{\dagger}a\rangle(A_{1} - B_{1} + A_{1}^{*} - B_{1}^{*}) + \langle a^{\dagger}b\rangle(C_{2}^{*} - D_{2})
\nonumber\\
&+&\langle b^{\dagger}a\rangle (C_{2} - D^{\ast}_{2}) + A_{1} + A_{1}^{*}, \nonumber\\
\frac{d}{dt}\langle b^{\dagger} b\rangle& = &\langle b^{\dagger} b\rangle(A_{2} - B_{2} + A_{2}^{*} - B_{2}^{*}) - \langle a^{\dagger}b\rangle(C_{1}^{*} - D_{1})
\nonumber\\
&-&\langle b^{\dagger}a\rangle (C_{1} - D^{\ast}_{1}) + A_{2} + A_{2}^{*},\nonumber\\
\frac{d}{dt}\langle a^{\dagger}b\rangle &=& \langle a^{\dagger}b\rangle\bigl(A^{\ast}_{1} - B_{1} + A_{2} - B^{\ast}_{2} - i(\Delta_{1}+ \omega)\bigr)
\nonumber\\
&-& \langle a^{\dagger}a\rangle(C_{1} - D^{\ast}_{1}) + \langle b^{\dagger}b\rangle (C_{2} - D^{\ast}_{2}) - C_{1} - D^{\ast}_{2}, \nonumber\\
\frac{d}{dt}\langle b^{\dagger}a\rangle &=& \langle b^{\dagger}a\rangle\bigl(A_{1} - B_{1}^{*} + A_{2}^{*} - B_{2} + i(\Delta_{1} + \omega)\bigr)
\nonumber\\
&-&\langle a^{\dagger} a\rangle(C_{1}^{*} - D_{1}) + \langle b^{\dagger} b\rangle (C_{2}^{*} - D_{2}) - C_{1}^{*} - D_{2}.  \nonumber \\ 
\label{ba}
\end{eqnarray}
\begin{figure}[t]
\centering
\includegraphics[height=5.3cm]{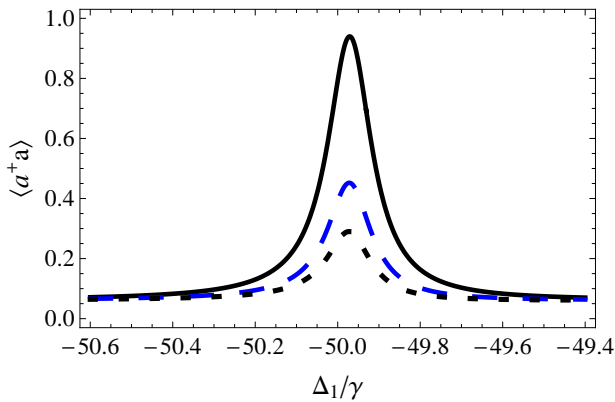}
\caption{\label{fig-2} 
(color online) The steady-state mean-value of the photon number $\langle a^{\dagger}a\rangle$ as a function of $\Delta_{1}/\gamma$. 
Here, $\gamma_{c}/\gamma=0.3$, $g/\gamma=2$, $\lambda/\gamma=4$, $\Omega/\gamma=50$, $\omega/\gamma=50$, 
$\Delta/(2\Omega)=0.5$, $\kappa_{a}/\gamma=0.01$ and $\kappa_{b}/\gamma=0.001$. The solid line corresponds to $\bar n=10$, 
the long-dashed one to $\bar n=4$ whereas the short-dashed line is for $\bar n=2$, respectively.}
\end{figure}
\begin{figure}[t]
\centering
\includegraphics[height=5.3cm]{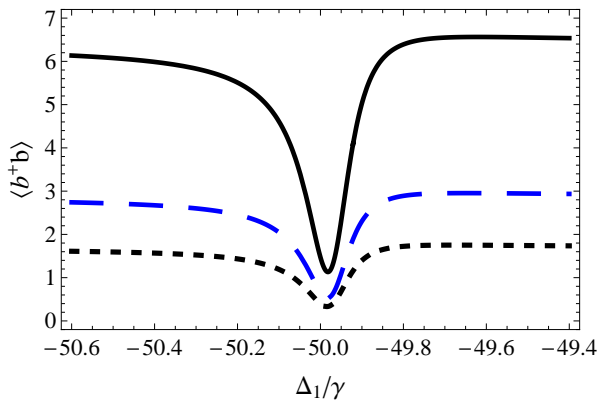}
\caption{\label{fig-3} 
(color online) The same as in Fig.~(\ref{fig-2}) but for the vibrational phonon mean-number $\langle b^{\dagger}b\rangle$.}
\end{figure}

Based on Eqs.~(\ref{ba}), Figures~(\ref{fig-2}) and (\ref{fig-3}) show the steady-states of the cavity mean-photon number, and the vibrational 
NMR mean-phonon number, respectively. As it was mentioned before, the maximum photon detection corresponds to NMR phonon minimum around 
$\Delta_{1}+\omega \approx 0$. 
Furthermore, the quantum cooling is still efficient while increasing the temperature, i.e. $\bar n$. These behaviors can be understood also by 
taking into account that for certain positive laser-qubit frequency detunings the qubit population is mostly in the lower dressed-state $|-\rangle$. 
This means that the phonon generation processes are minimized while phonon absorption followed simultaneously by photon laser absorption 
processes are accompanied by photon emission in the cavity mode. Therefore, the cavity maximum photon detection signifies also minimum of 
the vibrational quanta, that is, the cooling detection of the NMR vibrational degrees of freedom. Notice a small shift between photon maximum 
and phonon minimum detections for arbitrary larger qubit-cavity and qubit-NMR coupling strengths (see, also, Eq.~{\ref{meqm}}). Finally, efficient 
cooling occurs as well for uncorrelated regimes of phonon-photon detection processes, i.e. when $\Delta_{1}+ \omega \gg \gamma$ (while correlated 
regimes occur for $\Delta_{1} + \omega \ll \gamma$). However, in this case, i.e. uncorrelated regime,  it is hard to manage to have maximum photon 
cavity emission corresponding to vibrational NMR phonon cooling processes simply because these processes are uncorrelated. 

In what follows, we shall represent approximative analytical expressions for the variables of interest in the steady-state. This will help to 
understand the behaviors shown in Figs.~(\ref{fig-2}) and (\ref{fig-3}). If one performs further approximations 
$\{\Delta_{1},\omega\} \gg \Gamma_{\shortparallel}$, $2\Omega_{R} \pm \Delta_{1} \gg \Gamma_{\perp}$ and 
$2\Omega_{R} \pm \omega \gg \Gamma_{\perp}$, the master equation (\ref{MeCN}) simplifies considerably, namely,
\begin{eqnarray}
&&\frac{d}{dt}\rho(t)=\frac{i}{4}(\Delta_{1}+\omega-\bar \delta_{a} + \bar \delta_{b})[a^{\dagger}a-b^{\dagger}b,\rho] 
+ i\eta[ab^{\dagger},\rho] \nonumber \\
&&  - \kappa_{a}[a^{\dagger},a\rho] - \kappa_{b}(1+\bar{n})[b^{\dagger},b\rho] - \kappa_{b}\bar{n}[b,b^{\dagger}\rho] + H.c.. \nonumber \\
\label{meqm}
\end{eqnarray}
Here, the frequency shifts $\bar \delta_{a(b)}$ observed also in Fig.~(\ref{fig-2}) and Fig. (\ref{fig-3}) are given by the following expressions: 
$\bar \delta_{a}=g^{2}(P_{+}-P_{-})\{\sin^{4}{\theta}/(2\Omega_{R}+\omega) +  \cos^{4}{\theta}/(2\Omega_{R}-\omega)\}$ and $\bar \delta_{b}=\Omega_{R}\lambda^{2}\sin^{2}{2\theta}(P_{+}-P_{-})/(4\Omega^{2}_{R}-\omega^{2})$, whereas $\eta=g\lambda\sin{2\theta}(P_{+}-P_{-})
(\Omega_{R}\cos{2\theta}+\omega/2)/(4\Omega^{2}_{R}-\omega^{2})$. The last term of the first line in Eq.~(\ref{meqm}) with its H.c. part describe the vibrational 
phonon emission followed by cavity-photon absorption processes, and viceversa, mediated by the laser field. 
Consequently, basing on Eq.~(\ref{meqm}), Eqs.~(\ref{ba}) reduce to:
\begin{eqnarray}
\frac{d}{dt}\langle a^{\dagger}a\rangle &=& -i\eta \langle x\rangle - 2\kappa_{a}\langle a^{\dagger}a\rangle, \nonumber \\
\frac{d}{dt}\langle x\rangle &=& 2i\eta\bigl(\langle b^{\dagger}b\rangle - \langle a^{\dagger}a\rangle \bigr) - (\kappa_{a}+\kappa_{b})\langle x\rangle, \nonumber \\
\frac{d}{dt}\langle b^{\dagger}b\rangle &=& i\eta\langle x\rangle - 2\kappa_{b}\langle b^{\dagger}b\rangle + 2\kappa_{b}\bar n, \label{eqsm}
\end{eqnarray}
where $x=ab^{\dagger}-a^{\dagger}b$. We have assumed also that $\Delta_{1}+\omega = \bar \delta_{a} - \bar \delta_{b}$, i.e., 
we are interested in the maximal values of mean-photon number corresponding to minimal values of vibrational mean-phonon number, respectively
(see, also, Fig.~{\ref{fig-2}} and Fig.~{\ref{fig-3}}). In the steady-state, one immediately obtains from Eqs.~(\ref{eqsm}) that:
\begin{eqnarray}
\kappa_{a}\langle a^{\dagger}a\rangle + \kappa_{b}\langle b^{\dagger}b\rangle=\kappa_{b}\bar n. \label{eg}
\end{eqnarray}
This expression can help us to estimate the mean-vibrational-phonon number if the mean-photon number is known (i.e., detected). The explicit expressions for the 
steady-state values of the photon and phonon mean numbers are, respectively, 
\begin{eqnarray}
\langle a^{\dagger}a\rangle &=& \frac{\bar n\kappa_{b}\eta^{2}}{(\kappa_{a}+\kappa_{b})(\kappa_{a}\kappa_{b}+\eta^{2})}, \nonumber \\
\langle b^{\dagger}b\rangle &=& \frac{\bar n \kappa_{b}}{\kappa_{a}+\kappa_{b}}\biggl(1 + \frac{\kappa^{2}_{a}}{\kappa_{a}\kappa_{b}+\eta^{2}}\biggr ),
\label{ap_exp1}
\end{eqnarray}
or
\begin{eqnarray}
\langle b^{\dagger}b\rangle=\langle a^{\dagger}a\rangle\bigl(1 + \kappa_{a}(\kappa_{a}+\kappa_{b})/\eta^{2} \bigr). \label{ap_exp2}
\end{eqnarray}
Expressions (\ref{eg}) and (\ref{ap_exp2}) describe the efficiency of the proposed vibrational phonon cooling method. In particular, if 
$\kappa_{a} \gg \kappa_{b}$ while $(\kappa_{a}/\eta)^{2} \ll 1$, then 
$\langle b^{\dagger}b\rangle \approx \langle a^{\dagger}a\rangle \approx (\kappa_{b}/\kappa_{a})\bar n$ which can be as well below unity, i.e., 
$\langle b^{\dagger}b\rangle < 1$.
\section{Summary}
In summary, we have proposed a scheme to detect the vibrational phonon cooling of a nanomechanical oscillator in 
the steady-state. The idea is based on correlating the vibrational degrees of freedom with those of a laser-pumped 
quantum dot when fixed on a nanomechanical beam while interacting with an optical resonator. More concretely, when 
the quantum dot dynamics is faster than the corresponding ones of other involved subsystems,  one needs to adjust 
the laser frequency such that both photon laser and NMR phonon absorption processes are accompanied by photon 
emission in the resonator mode. Therefore, detection of the cavity photons is followed in parallel by cooling of the 
nanomechanical oscillator. Finally, we give approximative analytical expressions for the variables of interest which 
describe also the method efficiency.


\end{document}